\documentclass[conference]{IEEEtran}
\usepackage{graphicx}
\usepackage{subfigure}
\usepackage{multirow}
\usepackage{amssymb}
\usepackage{algorithmic}
\usepackage{bm}
\usepackage{mathrsfs}
\usepackage{threeparttable}
\usepackage{array}
\usepackage{indentfirst}
\makeatletter
\newif\if@restonecol
\makeatother

\usepackage[linesnumbered,ruled]{algorithm2e}
\usepackage{amsmath}
\usepackage{caption}
\usepackage{color}
\usepackage{url}
\usepackage{mathtools}
\usepackage{microtype}
\usepackage{adjustbox}
\usepackage{cite}

\begin{document}

\title{Joint Passive Beamforming and User Association Optimization for IRS-assisted mmWave Systems}

\author{\IEEEauthorblockN{Dan Zhao$^*$, Hancheng Lu$^*$, Yazheng Wang$^*$, Huan Sun$^\dagger$}
\IEEEauthorblockA{$^*$Key Laboratory of Wireless-Optical Communications, University of Science and Technology of China, Hefei, China.\\
$^\dagger$ Wireless Technology Laboratory, Huawei Technologies Co.,Ltd., Shanghai, China. \\
Email: $^*$zd2019@mail.ustc.edu.cn, $^*$hclu@ustc.edu.cn, $^*$wang1997@mail.ustc.edu.cn, $^\dagger$sunhuan11@huawei.com.}}

\maketitle

\begin{abstract}
In this paper, we investigate an intelligent reflect surface (IRS) assisted multi-user millimeter wave (mmWave) downlink communication system, exploiting IRS to alleviate the blockage effect and enhance the performance of the mmWave system. Considering the impact of IRS on user association, we formulate a sum rate maximization problem by jointly optimizing the passive beamforming at IRS and user association, which is an intractable non-convex problem. Then an alternating optimization algorithm is proposed to solve the problem efficiently. In the proposed algorithm, passive beamforming at IRS is optimized by utilizing the fractional programming method and user association is solved through the network optimization based auction algorithm.
We provide numerical comparisons between the proposed algorithm and different reference algorithms. Simulation results demonstrate that the proposed algorithm can achieve significant gains in the sum rate of all users.
\end{abstract}

\IEEEpeerreviewmaketitle

\section{Introduction}
The fifth generation (5G) mobile networks use key technologies such as millimeter wave (mmWave) technology, heterogeneous networks and massive multiple-input multiple-output (MIMO) to achieve 1000 times throughput increase and 10 times spectrum efficiency improvement\cite{2han2019achieving, Lu2015_Network}. One of the key performance indicators of 5G communication networks is high throughput.
As an emerging technology, mmWave communication can support a wide range of applications while meeting huge throughput requirements\cite{3busari2017millimeter}.
However, due to short wavelength of mmWave, the transmitted mmWave signals are difficult to pass through buildings, which can be easily absorbed. Hence, in order to conquer the block sensitive nature of mmWave and maintain robustness of the system, both the channel characteristics such as blockage effect and the line-of-sight (LoS)/non-line-of-sight (NLoS) propagation laws caused by such high frequency mmWave should be addressed.

Recently, \emph{intelligent reflect surfaces} (IRS) becomes a new promising technology to enhance the performance of wireless system. To address the problems above, we exploit the IRS to enhance the performance of mmWave communication system. IRS can intelligently reconfigure wireless propagation environment for directional signal enhancement or signal nulling\cite{11wu2019towards}.
Many efforts have been dedicated to the research of IRS\cite{13guo2019weighted, 14cao2019intelligent, 15han2019large, 17wu2018intelligent}, such as increasing sum rate of all users \cite{13guo2019weighted} and enhancing communication coverage \cite{14cao2019intelligent}.
In particular, IRS resembles a full-duplex \emph{amplify-and-forward} (AF) relay, and it consists of a large number of passive reflective elements, which can forward incident signals using passive beamforming.
Besides, IRS is capable of adaptively shaping the propagation channels based on actual environmental conditions by inducing certain phase shift to incident signals. Therefore, the hardware cost and power consumption of IRS are far lower than those of AF relay.

Although many benefits can be brought by IRS to mmWave system, it is worth noticing that IRS has a significant impact on
existing user association algorithms \cite{5liu2016user, 8yang2017association, 9athanasiou2014optimizing}.
In mmWave communication systems, the received signal strength based user association algorithm is widely used \cite{5liu2016user}, which may lead to inefficient use of resources.
There are some research efforts to improve the throughput of system by considering factors such as load balancing and fairness in mmWave communication systems \cite{9athanasiou2014optimizing}. The study in \cite{8yang2017association} aims to maximize network utility by jointly optimizing user association, load distribution and power control.
In IRS-assisted mmWave communication systems, the impact on channels involved by IRS should be considered. IRS has great potential to enhance the performance of mmWave communications, especially in combination with user association. Thus, we need to jointly consider IRS and user association to further improve the performance of the mmWave communication systems. 

In this paper, we study the joint IRS and user association optimization problem in IRS-assisted mmWave systems. Compared with the mmWave system leveraging AF relay, the scenario we consider has lower hardware cost and energy consumption.
However, the integration of IRS and user association has great challenges in mmWave communication systems.
Existing works on IRS mostly focuses on system performance optimization in a single base station (BS) scenario, which cannot be adopted in the scenario of multiple BSs, and cannot guarantee the robustness of the entire mmWave system.
Moreover, after introducing IRS in a multi-BS scenario, the passive beamforming at IRS couples with user association due to its impact on channels,
so we have to redesign an effective user association algorithm.
Furthermore, the mmWave channel characteristics make the optimization of user association much more complicated than those of existing works \cite{8yang2017association, 9athanasiou2014optimizing}.

We consider an IRS-assisted multi-user mmWave downlink system, where an IRS with discrete phase shifter is deployed to assist the downlink transmission. By optimizing the user association, each BS can simultaneously serve multiple users. Simultaneously, we utilize IRS to reflect transmitted signal and deal with the impact of blockages. Here, we aim to maximize the sum rate of all users under the constraints of discrete phase shift for IRS and maximum transmit power at BSs via jointly optimizing the passive beamforming at IRS and user association. Given the non-convexity and complexity of the problem, we utilize the alternating optimization algorithm. Specifically, the passive beamforming at IRS is optimized based on fractional programming method, and the user association problem is solved through the auction algorithm and network optimization. Simulation results show that the proposed algorithm can maximize the sum rate of the system significantly.

The rest of this paper is organized as follows. In Section II we describe the system model and formulate a sum rate maximization problem. In Section III, an algorithm based on the alternating optimization is designed to solve the problem. Simulation results of the proposed algorithm are discussed in Section IV. Finally, a conclusion is given for this paper in Section V.
\vspace{-1.0em}
\section{System Model and Problem Formulation}
\subsection{System Model}
\begin{figure}[htbp]
  \centering
  \includegraphics[width=7.5cm]{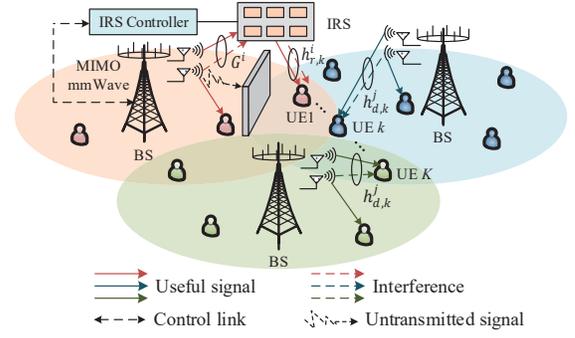}\\
  \caption{An IRS-assisted multi-user mmWave system.}
  \label{fig:IRS}
  \vspace{-1.5em}
\end{figure}

As shown in Fig.\ref{fig:IRS}, we consider an IRS-assisted downlink multi-user mmWave multiple input single output system with $K$ users served by $S$ BSs. An IRS with $N$ reflect elements is deployed on the surface of the building and
reflects the transmitted signal from the assisted BS for its associated users. 
Each BS is equipped with $M>K$ antennas, and each user is equipped with only one antenna. In downlink transmission, BS can transmit independent data steams to these $K$ users simultaneously. For convenience, the set of the user indexes and the BS indexes are denoted as $\mathcal{K}=\{1,2,\cdots,K\}$ and $\mathcal{S}=\{1,2,\cdots,S\}$, respectively. Since the direct links between the IRS-assisted BS $s\in\mathcal{S}$ and users are severely blocked by obstacles, we suppose that user $k\in\mathcal{K}$ communicates with BS $s$ through the IRS reflection link. Furthermore, we assume that only BS $i\in \mathcal{S}$ is assisted by IRS and other BSs are not equipped with IRS. $\bm{G}^i \in \mathbb{C}^{N\times M}, \bm{h}^i_{r,k} \in \mathbb{C}^{1\times N}, \bm{h}^j_{d,k} \in \mathbb{C}^{1\times M}, j\in \mathcal{S}, i\neq j$ are equivalent channels from the BS $i$ to the IRS, from the IRS to all users, and from the BS $j$ to all users, respectively. Besides, let $\bm{H}_k=\left[\bm{h}^{1H}_{k},\cdots, \bm{h}^{SH}_{k}\right] \in \mathbb{C}^{M\times S}$ denote the channels between user $k$ and all BSs, and $\bm{h}^s_k \in \{\bm{h}^i_{r,k}\boldsymbol{\Phi}\bm{G}^i, \bm{h}^j_{d,k}\}$ denotes the channel between user $k$ and BS $s$. The phase shift matrix of IRS is denoted as $\boldsymbol{\Phi} =  diag\{\beta_1 e^{j\varphi_{1}},\beta_2 e^{j\varphi_{2}},\cdots,\beta_N e^{j\varphi_{N}}\}$, where $\varphi_{n}$ and $\beta_n$ represent the phase shift and the amplitude reflection coefficient of the $n$-th element of the IRS, respectively. In practice, each element of the IRS is usually designed to maximize signal reflection (\emph{i.e.,} $\beta_n$ = 1).

We define the user association vector $\bm{A} \in \mathbb{C}^{S\times K}$ in (\ref{12}), where $\boldsymbol{\tilde{a}}_s$ and ${{\left\| {{{\boldsymbol{\tilde{a}}}}_{s}} \right\|}_{0}}$ denote the user association vector at the BS $s$ and the number of users served by $s$, respectively, and $\bm{a}_k={{\left[a_{1,k},a_{2,k},...,a_{S,k} \right]}}^T\in\mathbb{C}^{S\times 1}$ represents the association vector of user $k$. Since each user is served by only one BS, ${{\left\| {{{\boldsymbol{a}}}_{k}} \right\|}_{0}}=1$ and $a_{s,k}$ is a binary variable (\emph{i.e.,} $a_{s,k} \in \{0, 1\}$).
\begin{equation}\label{12}
\resizebox{0.9\hsize}{!}{$
\bm{A} = \left[\bm{a}_1, \bm{a}_2,\cdots , \bm{a}_K \right]=\left[
\begin{array}{cccc}
\underbrace{a_{11}, a_{12}, \cdots, a_{1K}}_{\widetilde{\bm{a}}_1}\\
 \vdots \\
 \underbrace{a_{S1}, a_{S2}, \cdots, a_{SK}}_{\widetilde{\bm{a}}_S}
\end{array}
\right],$}
\end{equation}

In particular, we need to determine BS $s \in \mathcal{S}$ that communicates with user $k$ and the channel that user $k$ actually transmits information $\bm{h}^s_k \in \mathbb{C}^{1\times M}$ via user association vector $\bm{a}_k$.
The data symbol $t^s_k$ transmitted from BS $s$ to user $k$ is an independent variable with zero mean and unit power, and $\boldsymbol{\omega}^s_k \in \mathbb{C}^{M\times 1}$ is the corresponding beamforming vector at BS $s$. The signal received by user $k$ from BS $s$ can be expressed as

\begin{equation}\label{16}
y_{s,k} = \bm{h}^s_k \bm{\omega}^s_k t^s_k + \bm{h}^s_k \sum_{j\neq k, j \in Q_s} \bm{\omega}^s_j t^s_j +n_k, k\in \mathcal{K},
\end{equation}
where $Q_s$ represents the index set of users associated with BS $s$, $n_k \sim \mathcal{CN}(0,\sigma^2)$ is complex additive white Gaussian noise with zero mean and variance $\sigma^2$. We assume that neighboring BSs can be allocated orthogonal frequency band or employ enhanced inter-cell interference coordination techniques \cite{Guoastely2009lte}. Accordingly, the signal-to-interference-plus-noise ratio (SINR) of user $k$ is written as $\gamma_k$ and it can be given by
\begin{equation}\label{SNRk}
\gamma_{k} = \frac{\sum_{s=1}^{S} a_{s, k} \left| \bm{h}^s_k \bm{\omega}^s_k \right|^2}{\sum_{j\neq k, j \in Q_s}\left| \bm{h}^s_k \bm{\omega}^s_j \right|^2 + \sigma^2}, \forall k, \forall s.
\end{equation}

Suppose that the bandwidth $B$ allocated to each BS is equal, and the achievable rate from BS $s$ to user $k$ is defined as $R_{s,k}$ which can be computed as follows
\begin{equation}\label{18}
R_{s, k} = B \log_2\left(1 + \frac{\left| \bm{h}^s_k \bm{\omega}^s_k \right|^2}{ \sum_{j\neq k, j \in Q_s}\left| \bm{h}^s_k \bm{\omega}^s_j \right|^2 + \sigma^2}\right).
\end{equation}

\subsection{Channel Model}
To characterize the theoretical performance gain brought by IRS, we assume that the channel state information of all channels involved is perfectly known at the BS. For the IRS-assisted BS $i$, we only consider the BS-IRS channel and IRS-user channels. The BS-IRS channel is modeled according to the geometric channel model \cite{14cao2019intelligent} as
\begin{equation}\label{4}
\bm{G}^i = \sum_{g=0}^{G_p}\alpha_g \xi_t \xi_r \bm{a}^H_N \left(\theta_{AoA}^{(g)} \right) \bm{a}_M \left(\theta_{AoD}^{(g)} \right),
\end{equation}
where $G_p$ denotes the number of NLoS paths and $g = 0$ denotes the LoS path, $\alpha_g$ indicates the complex gain of the $g$-th path, and $\xi_r$ and $\xi_t$ are the receive and transmit antenna gains, respectively.
We employ a uniform linear array (ULA) at the IRS \cite{15han2019large}. The parameters $\theta_{AoA}^{(g)}$ and $\theta_{AoD}^{(g)}$ represent the angle of arrival (AoA) and the angle of departure (AoD) of the signal reflected by the IRS in the $g$-th path. Thus, the array response of the IRS $\bm{a}_N (\theta_{AoA})$ can be expressed as
\begin{equation}\label{11}
\bm{a}_N(\theta_{AoA}) = \left[1, e^{j2\pi\frac{d}{\lambda} sin\theta_{AoA}},\cdots, e^{j2\pi\frac{d}{\lambda} (N - 1)sin\theta_{AoA}}   \right]\!.\!\!
\end{equation}

Assume that the IRS is coated on the buildings around users, then the channel between the IRS and $k$-th user can be described as
\vspace{-0.5em}
\begin{equation}\label{zd}
\bm{h}^i_{r, k} = \alpha_k \xi_t \xi_r \bm{a}_N \left(\theta_{AoD} \right),
\end{equation}
where $\alpha_k$ indicates the complex gain and $\theta_{AoD}$ is the AoD of the signal from the IRS to the $k$-th user. Besides, BS-user channels can be also obtained similar to (\ref{zd}).
\subsection{Problem Formulation}
We further define the channel and beamforming matrix of BS $s$ as $\bm{H}^H_s=[\bm{h}^{sH}_{1},\cdots, \bm{h}^{sH}_{\|Q_s\|}] \in \mathbb{C}^{M\times \|Q_s\|}$ and $\bm{W}_s=[{\boldsymbol{\omega}^s_{1}},\cdots,{\boldsymbol{\omega}^s_{\|Q_s\|}}] \in \mathbb{C}^{M\times \|Q_s\|}$, respectively. In this paper, our goal is to maximize the sum rate of all users by jointly optimizing the passive reflect beamforming and user association in an IRS-assisted mmWave system under constraints.
Accordingly, the sum rate maximization problem is formulated as
\begin{subequations} \label{p1}
\begin{align}
\rm{P1}: \mathop{\max}_{\boldsymbol{\Phi},\boldsymbol{A}}~~&
f_1(\bm{\Phi}, \bm{A}) = B \sum^K_{k = 1}\log_2\left(1+ \gamma_{k}\right),\label{f1}\\
s.t.~~& \varphi_n \in \mathcal{F}, \forall n, \label{251}\\
& \sum_{s \in \mathcal{S}} a_{s, k} = 1, \forall k, \label{22}\\
& \sum_{k \in \mathcal{K}} a_{s, k} \geq 1, \forall s, \label{23}\\
& \!\sum_{k \in Q_s} \left\|\bm{\omega}^s_k\right\|^2 \leq P_s, \forall s, \label{24}
\end{align}
\end{subequations}
where $\mathcal{F}=\left\{ 0,\frac{2\pi }{{{2}^{b}}},\frac{2\pi \times 2}{{{2}^{b}}},\cdots,\frac{2\pi \times ({{2}^{b}}-1)}{{{2}^{b}}} \right\}$ is the set of available phase shifts for the IRS, $b$ is the resolution of the phase shifter at IRS and $P_s$ is the tranmit power of BS $s$. Also, constraint (\ref{22}) ensures that each user communicates with only one BS, constraint (\ref{23}) indicates that each BS can serve multiple users simultaneously, while constraint (\ref{24}) accounts for the fact that the transmit power of BS $s$ is kept below the maximum threshold $P_s$. As there are more antennas than users, simple linear processing techniques such as maximum ratio transmission (MRT) or zero forcing (ZF) are near optimal \cite{van2016joint}. Therefore, we consider using ZF precoding at each BS and satisfy constraint (\ref{24}) with $\bm{W}_{s}^{opt}=\sqrt{P_s}\frac{\bm{H}_s^H}{\| \bm{H}_s\bm{H}_s^H\|}$.
However, due to the non-convex objective function (\ref{f1}) and the non-convex constraint (\ref{251}), problem P1 cannot be solved globally. In the next section, we propose an algorithm exploiting alternative optimization algorithm to solve the problem P1.

\section{Proposed Solution}
In this section, we exploit alternating optimization algorithm to optimize $\bm{\Phi}$ and $\bm{A}$. For given $\bm{A}$, we apply the Lagrangian dual transform to decouple $f_1({\bm{\Phi},\bm{A}})$, and optimize $\bm{\Phi}$ based on the fractional programming method. Then for given $\bm{\Phi}$, we consider the network optimization and design a more efficient algorithm in combination with the auction algorithm to optimize $\bm{A}$.
\subsection{IRS Reflection Matrix Optimization}
This subsection shows how to optimize IRS reflection matrix $\bm{\Phi}$ with given user association vector. When the user association vector $\bm{A}$ is fixed, problem P1 can be equivalently written as
\vspace{-0.5em}
\begin{equation}\label{p2}
\begin{split}
\rm{P2}: \mathop{\max}_{\boldsymbol{\Phi}} ~~&
f_2(\bm{\Phi}) = B \sum^K_{k = 1}\log_2\left(1+ \gamma_{k}\right), \\
s.t.~~& \varphi_n \in \mathcal{F}, \forall n.
\end{split}
\end{equation}
The Lagrangian dual transform proposed in \cite{22shen2018fractional} is used to tackle the logarithm in the objective function of P2. By introducing an auxiliary variable
$\boldsymbol{\lambda}=\left[\lambda_1,\lambda_2,\cdots,\lambda_k\right]^T$, the objective function can be represented as
\begin{equation}
\resizebox{0.88\hsize}{!}{$\!\!\!f_{3}(\boldsymbol{\Phi},\! \boldsymbol{\lambda})\!\! =\!\! B\!\!\left(\displaystyle\sum_{k=1}^{K}\log_2(1\!\!+\!\!\lambda_k)\!\!-\!\!\sum_{k=1}^{K}\lambda_k \\
\!\!+\!\!\sum_{k=1}^{K}\!\!\frac{(1\!\!+\!\!\lambda_k)\gamma_k}{1\!\!+\!\!\gamma_k}\!\!\right)\!\!.$}
\end{equation}
When $\boldsymbol{\Phi}$ is fixed, the optimal $\lambda_k$ is $\lambda_k^{opt}=\gamma_k$. Then for a fixed $\boldsymbol{\lambda}$, the problem can be transformed as
\begin{equation}\label{p3}
\begin{split}
\underset{\boldsymbol{\Phi}}{\mathop{\max}}\, ~~& f_4(\boldsymbol{\Phi})=\sum_{k=1}^{K}\frac{(1+\lambda_k)\gamma_k}{1+\gamma_k}, \\
s.t.~~& \varphi_n \in \mathcal{F}, \forall n.
\end{split}
\end{equation}

Given the user association vector, the set of users served by the BS $i$ is symbolized as $\mathcal{M}=\{1,2,\cdots,M\}$. Using the expression of $\gamma_k$ in (\ref{SNRk}), $f_4(\boldsymbol{\Phi})$ can be expressed as
\vspace{-0.5em}
\begin{equation}\label{f4}
\resizebox{0.7\hsize}{!}{$
\begin{split}
f_4(\boldsymbol{\Phi})& = \sum^{M}_{m = 1} \frac{(1+\lambda_{m}) \left|\bm{h}_{m}^i \bm{\omega}^i_{m}\right|^2}{ \sum_{j= 1}^{M} \left|\bm{h}_{m}^i \bm{\omega}^i_j\right|^2 + \sigma^2}\\
& = \sum^{M}_{m = 1} \frac{(1+\lambda_{m}) \left|\bm{h}_{r, m}^i \bm{\Phi}\bm{G}^i \bm{\omega}^i_m \right|^2}{ \sum_{j= 1}^{M} \left|\bm{h}_{r, m}^i \bm{\Phi} \bm{G}^i \bm{\omega}^i_j\right|^2 + \sigma^2}.
\end{split}
$}
\end{equation}

Since $\boldsymbol{\Phi}$ is a diagonal matrix, we define
$\boldsymbol{b}_{m,j}=diag(\bm{h}_{r, m}^i)\bm{G}^i\boldsymbol{\omega}^i_j$, $\boldsymbol{\varphi}=[e^{j\varphi_{1}},\cdots,e^{j\varphi_{N}}]^T=[\theta_1,\cdots,\theta_n]^T$. Then (\ref{p3}) can be equivalently transformed as
\begin{equation}\label{p4}
\resizebox{0.7\hsize}{!}{$
\begin{split}
\underset{\bm{\varphi}}{\mathop{\max}}\, ~~& f_5(\bm{\varphi})=\sum_{m=1}^{M}\frac{(1+\lambda_m)\left|\bm{\varphi}^H\bm{b}_{m,m}\right|^2}{\sum_{j= 1}^{M}
\left|\bm{\varphi}^H\bm{b}_{m,j}\right|^2+\sigma^2}, \\
s.t.~~& \varphi_n \in \mathcal{F}, \forall n.
\end{split}
$}
\end{equation}

Note that (\ref{p4}) is a multiple-ratio fractional programming problem, which can be transformed to the following problem based on the quadratic transform proposed in \cite{22shen2018fractional}
\begin{equation}\label{p5}
\begin{split}
\underset{\bm{\varphi},y}{\mathop{\max}}\, ~~& f_6(\bm{\varphi},y),\\
s.t.~~& \varphi_n \in \mathcal{F}, \forall n,
\end{split}
\end{equation}
where the objective function is
\begin{equation}\label{f6}
\begin{split}
f_6(\bm{\varphi},y)=& \sum^{M}_{m = 1} 2\sqrt{1+\lambda_m}\operatorname{Re}\left\{y^*_m \bm{\varphi}^H \bm{b}_{m, m}\right\}\\
& - \sum^{M}_{m = 1} |y_m|^2 \left(\left|\bm{\varphi}^H \bm{b}_{m, j}\right|^2 + \sigma^2\right),
\end{split}
\end{equation}
and $\bm{y}=[y_1,y_2,\cdots,y_K]^{T}$ is an auxiliary variable vector.

Similarly, problem (\ref{p5}) can be solved by alternatively optimizing $\bm{y}$ and $\bm{\varphi}$. With given $\bm{\varphi}$, the optimal $y_m$ can be obtained by setting $\partial f_6/\partial {y_m}$ to zero as
\begin{equation}\label{yopt}
y^{opt}_m = \frac{\sqrt{1+\lambda_m}\bm{\varphi^H}\bm{b}_{m, m}}{\sum^{M}_{j = 1}\left|\bm{\varphi}^H \bm{b}_{m, j}\right|^2 + \sigma^2}.
\end{equation}
Then the remaining problem is to optimize $\bm{\varphi}$ for a given $\bm{y}$. Using the relationship that $\left|\bm{\varphi}^H \bm{b}_{m, j} \right|^2=\bm{\varphi}^H \bm{b}_{m, j}\bm{b}^H_{m,j}\bm{\varphi}$, the optimization problem for $\bm{\varphi}$ can be represented as follows
\begin{equation}\label{p6}
\begin{split}
\underset{\bm{\varphi}}{\mathop{\max}}\, ~~& f_7\left(\bm{\varphi}\right)=-\bm{\varphi}^H\bm{U}\bm{\varphi}+2\operatorname{Re}\{\bm{\varphi}^H\bm{v}\}+C, \\
s.t.~~& \varphi_n \in \mathcal{F}, \forall n,
\end{split}
\end{equation}
where
\begin{subequations}\label{uvc}
\begin{align}
& \bm{U}=\sum_{m=1}^{M}\left|y_m\right|^2 \sum^{M}_{j = 1}\bm{b}_{m,j}\bm{b}_{m,j}^H, \label{u}\\
& \bm{v}=\sum_{m=1}^{M}\sqrt{1+\lambda_m}y^*_m\bm{b}_{m,m}, \label{v}\\
& C=-\sum_{m=1}^{M}\left|y_m\right|^2\sigma^2. \label{c}
\end{align}
\end{subequations}

Similar to \cite{13guo2019weighted}, we iteratively optimize one of the element of $\bm{\varphi}$ by keeping the other $N-1$ phases fixed. Denote the element at $i$-th row and $j$-th column of $\bm{U}$ by $u_{i,j}$, and the $i$-th element of $\bm{v}$ by $v_i$. Since $\bm{U}$ is a Hermitian matrix, we can represent $\bm{\varphi}^H \bm{U} \bm{\varphi}$ and $\bm{\varphi}^H \bm{v}$ as
\begin{equation}\label{30}
\begin{split}
\bm{\varphi}^H \bm{U} \bm{\varphi} = ~& \theta^*_n u_{n, n} \theta_n + \sum^N_{i = 1, i \neq n} \sum^{N}_{j = 1, j \neq n} \theta^*_i u_{i, j} \theta_j\\
~& + 2\operatorname{Re}\{\sum^N_{j = 1, j \neq n} \theta^*_n u_{n, j} \theta_j\},
\end{split}
\end{equation}
\begin{equation}\label{31}
\bm{\varphi}^H \bm{v} =\theta^*_n v_n + \sum^{N}_{i = 1, i \neq n} \theta^*_i v_i.
\end{equation}

Substituting (\ref{30}) and (\ref{31}) into (\ref{p6}) and dropping all the irrelevant constants, we can rewrite $f_7(\bm{\varphi})$ as
\begin{equation}\label{f8}
f_8\left(\theta_n\right)=-|\theta_n|^2 u_{n, n}+2\operatorname{Re}\{\theta^*_n(v_n - \sum^N_{j = 1, j \neq n} u_{n, j} \theta_j)\}.
\end{equation}
We define $d_n=v_n - \sum^N_{j = 1, j \neq n} u_{n, j} \theta_j$ and denote the argument of $d_n$, $\theta_n$ by $\angle d_n$, $\angle \theta_n$ (\emph{i.e.,} the phase shift of reflect element on IRS). Thus, the optimal solution $\theta_n$ can be obtained as
\begin{equation}\label{opttheta}
\angle \theta_{n}^{opt}=\arg \underset{\angle \theta_n \in \mathcal{F}}{\mathop{\min }}\,\left| \angle \theta_n -\angle d_n \right|.
\end{equation}
Finally, all the reflection coefficients from $n = 1$ to $n = N$ can be repeatedly optimized based on (\ref{f8}).

\subsection{User Association Optimization}
In this subsection, by applying the network optimization structure, an auction-based algorithm is proposed to optimize the user association vector $\bm{A}$ with given $\bm{\Phi}$. We denote the set of users that BS $s$ can serve as $\mathcal{A}(s)$ and the set of BSs that can serve user $k$ as $\mathcal{B}(k)$. Moreover, an assignment $\mathcal{C}$ is used to indicate a set of BS-user pairs $(s,k)$ with $k \in \mathcal{A}(s)$. Therefore, the association problem is reformulated by the following linear optimization problem as
\vspace{-0.5em}
\begin{subequations} \label{p9}
\begin{align}
\rm{P3}:\mathop{\max}_{\bm{A}} ~~& \sum_{s \in \mathcal{S}}\sum_{k \in \mathcal{K}} R_{s, k} a_{s, k},
\label{39}\\
s.t.~~& \!\!\!\sum_{k \in \mathcal{A}(s)} a_{s, k} \geq 1, \forall s, \label{40}\\
& \!\!\!\sum_{s \in \mathcal{B}(k)} a_{s, k} = 1, \forall k. \label{41}
\end{align}
\end{subequations}

As the structure of problem P3 is consistent with the typical minimum cost flow problem \cite{6athanasiou2013auction}, we convert problem P3 into a typical minimum cost flow problem by introducing a virtual node $e$ connected to each BS as
\vspace{-0.5em}
\begin{subequations}\label{p10}
\begin{align}
\rm{P4}:\mathop{\min}_{\bm{A}} ~~& \sum_{s \in \mathcal{S}} \sum_{k \in \mathcal{K}} -R_{s, k} a_{s, k},\\
s.t. ~~& \!\!\!\!\sum_{k \in \mathcal{A}(s)} a_{s, k} - a_{e, s} = 1, \forall s, \label{43}\\
& \!\sum_{s \in \mathcal{S}} a_{e, s} = K - S, \label{44}\\
& \!\!\!\!\!\sum_{s \in \mathcal{B}(k)} a_{s, k} = 1, \forall k, \label{45}
\end{align}
\end{subequations}
where $a_{s,k}$ means the amount of flow between BS $s$ and user $k$. Constraint (\ref{43}) ensures that the flow supply of each BS $s$ is one unit, constraint (\ref{44}) declares that $e$ is the source node with $K-S$ units of flows, and constraint (\ref{45}) ensures that each user is served by only one BS.
Besides, we utilize the duality theory to rewrite the minimum cost flow problem to
\vspace{-0.5em}
\begin{subequations}\label{p11}
\begin{align}
\mathop{\min}_{\pi_s, p_k, \mu} ~~& \sum_{s \in \mathcal{S}} \pi_s + \sum_{k \in \mathcal{K}} p_k + (K - S)\mu \label{46},\\
s.t. ~~&\pi_s + p_k \geq R_{s, k}, \forall s, \forall k, \label{47} \\
&\mu \geq \pi_s, \forall s \label{48},
\end{align}
\end{subequations}
Here, $\pi_s$, $\mu$ and $p_k$ are all Lagrangian multipliers and they are associated with constraint (\ref{43}), (\ref{44}) and (\ref{45}), respectively. The parameter $-\pi_s$ represents the price of each BS $s$, $\mu$ stands for the price of source node $e$ and $p_k$ denotes the  price of each user $k$.
We introduce $\epsilon-Complementary \ Slackness$ ($\epsilon-CS$) to solve problem (\ref{p11}). Let $\epsilon$ be a positive scalar. Then an assignment $\mathcal{C}$ and a pair $(\pi,p)$ satisfy $\epsilon-CS$ if
\begin{subequations}\label{eee}
\begin{align}
&\pi_s + p_k \geq R_{s, k} - \epsilon, \forall s, \forall k, \label{49}\\
&\pi_s + p_k = R_{s, k}, \forall(s, k) \in \mathcal{C}, \label{50}\\
&\pi_s = \mathop{\max}_{l = 1, 2, \cdots, S} \pi_l, \forall s. \label{51}
\end{align}
\end{subequations}

\begin{algorithm}[t]
\fontsize{9.5pt}{9.5pt}\selectfont
\caption{\mbox{Auction-based User Association Algorithm}} \label{Auction}
\SetKwInOut{KwIn}{Require}
\SetKwInOut{KwOu}{Ensure}
\nlset{1}\KwIn{Initial values of $\mathcal{C},(\pi,p), \epsilon$ and $\mu$}
\nlset{2}\KwOu{$R_{s,k}-p_k \geq \max\limits_{l \in \mathcal{A}(s)}\{R_{s,k}-p_l\} - \epsilon, \forall (s, k) \in \mathcal{C}$}
    \While{there are unassigned BSs}
    {
        BS $s$ is unassigned in $\mathcal{C}$, find the best user $k_s$ that: \mbox{$k_s = \mathop{\arg\max}\limits_{k \in \mathcal{A}(s)} \left\{R_{s,k} - p_k\right\}$, $ \rho_s = \max\limits_{k \in \mathcal{A}(s)} \left\{R_{s,k} - p_k\right\}$}, $\omega_s = \max_{k \in \mathcal{A}(s), k \neq k_s} \left\{R_{s,k} - p_k\right\}$;\\
        \If{$k_s$ is the only user in $\mathcal{A}(s)$}
        {
            $\omega_s \rightarrow -\infty$;\\
        }
        $ b_{s,k_s} = p_{k_s} + \rho_s - \omega_s + \epsilon = R_{s,k_s} - \omega_{s} + \epsilon$;\\
        $p_k = \max\limits_{s \in P(k)} b_{s,k}$, where $P(k)$ is the set of BSs that user $ k $ received a bid;\\
        Remove any pair $(s, k)$, where $ k $ was initially assigned to some $ s $ under $ \mathcal{C} $, and add the pair $(s_k, k)$ to $ \mathcal{C} $ with $ s_k = \mathop{\arg\max}_{s \in P(k)} b_{s,k} $.\\
    }
    \For{all unassociated users}
    {
        User $ k $ is unassociated in $ \mathcal{C} $, find the best BS $ s_k $ that: \mbox{$s_k = \mathop{\arg\max}\limits_{s \in \mathcal{B}(k)} \left\{R_{s,k} - \pi_s\right\}$, $ \zeta_k = \max\limits_{s \in \mathcal{B}(k)} \left\{R_{s,k} - \pi_s\right\}$}, $\omega_k = \max_{s \in \mathcal{B}(k), s \neq s_k} \left\{R_{s,k}- \pi_s\right\}$;\\
            \If{$s_k$ is the only BS in $\mathcal{B}(k)$}
    {
        $\omega_k \rightarrow -\infty$;\\
    }
        $ \delta = \min\left\{\mu - \pi_{s_k}, \zeta_k - \omega_k + \epsilon \right\}$;\\
        add $(s_k, k)$ to $ \mathcal{C}$: $ p_k = \zeta_k - \delta, \pi_{s_k} = \pi_{s_k} + \delta $ \\
        \If{$\delta > 0$}
        {
            Remove the pair $ (s_k, k_{\text{old}}) $ where $k_{\text{old}}$ was initially assigned to $ s_k $ under $ \mathcal{C} $;\\
        }
    }

    \KwOut{A feasible assignment $\mathcal{C}$.}
\end{algorithm}
Considering a dual variable pair $(\pi, p)$ and let $\mathcal{C}$ be a feasible solution for problem P3. Assuming that $\epsilon < 1/S$ and $R_{sk}/R_{min}$ is an integer $\forall {s, k}$, $\mathcal{C}$ is the optimal solution of problem P3, if $\epsilon-CS$ conditions (\ref{eee}) are satisfied by $\mathcal{C}$ and $(\pi, p)$ \cite{6athanasiou2013auction}. Based on the above theory, problem (\ref{p11}) can be solved by auction-based algorithm, as shown in Algorithm \ref{Auction}. Algorithm \ref{Auction} first associates each BS with one user (line 3-11), next assigns the remaining users to available BSs (line 12-22), and then obtains the optimal solution.

Finally, we solve the problems P2 and P3 by iterating continuously and alternately, and make the problem P1 reach a stable optimal rate.

\begin{figure}[t]
  \centering
  \includegraphics[width=5.6cm]{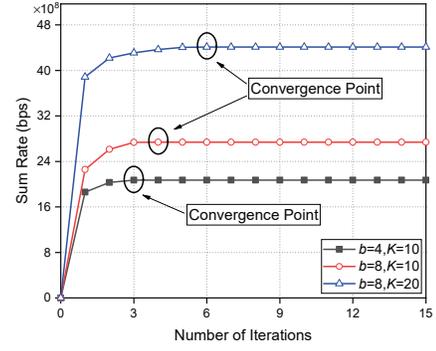}
  \caption{Convergence performance of the proposed algorithm.}
  \label{fig:Iteration}
\end{figure}
\vspace{-0.5em}
\begin{figure}[t]
  \centering
  \includegraphics[width=5.6cm]{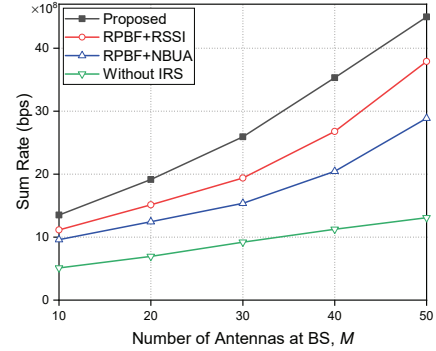}
  \caption{Sum rate versus $M$ with $N=60$.}
  \label{fig:M}
\end{figure}
\section{Simulation Results}
We consider the IRS comprising $N=60$ passive elements arranged in an ULA and $10$ users are randomly distributed in a circle at $(200\ m,0\ m)$ with radius $50\ m$. There are two BSs located at $(0\ m,0\ m)$ and $(400\ m,0\ m)$, respectively, and each of them is equipped with $30$ antennas.
For all the simulations, we assume the IRS-assisted multi-user mmWave system operates at 28\ GHz with bandwidth $B=100$ MHz.
The complex gain $\alpha_k$ is generated according to a
complex Gaussian distribution $\alpha_k \sim \mathcal{CN}(0,10^{-0.1\kappa})$ and $\kappa=\kappa_a+10\kappa_b \log_{10}(d)+\kappa_c$ with $\kappa_a=72$, $\kappa_b=2.92$, $\kappa_c \sim \mathcal{N}(0,\sigma_c^2)$ and $\sigma_c=8.7$ dB, and the generation of $\alpha_g$ is similar to $\alpha_k$.
Other required parameters are set as follows: $b=8$, $\sigma^2=-117$ dBm, $P_s=30$ dBm, $G_p=5$, $\xi_t=9.82$ dBi, $\xi_r=0$ dBi, $\epsilon=0.2$.

In this paper, three algorithms are considered to compare the performance with our proposed algorithm: 1) \text{RPBF + RSSI}: uses the received signal strength indicator (RSSI) based user association algorithm \cite{5liu2016user} and the random passive beamforming (RPBF) is applied at IRS; 2) \text{RPBF + NBUA}: uses the nearest-based user association (NBUA) \cite{zuo2016energy} algorithm and the RPBF is applied at IRS; 3) \text{Without IRS}: considers the sum rate of 10 users served by two BSs and each of them uses ZF precoding to serve 5 users without IRS (\emph{i.e.,} $N=0$) and user association optimization, which is plotted as a benchmark.

Fig. \ref{fig:Iteration} presents the convergence performance of the proposed algorithm, under different number of users and phase resolutions at IRS. We can see that the minimum $b$ and $K$ converges fastest and all these three curves converge to stable solutions after no more than 6 iterations. With the increase of $b$ and $K$, the convergence speed becomes slower but obtains significant improvement on sum rate.
\begin{figure}[t]
  \centering
  \includegraphics[width=5.6cm]{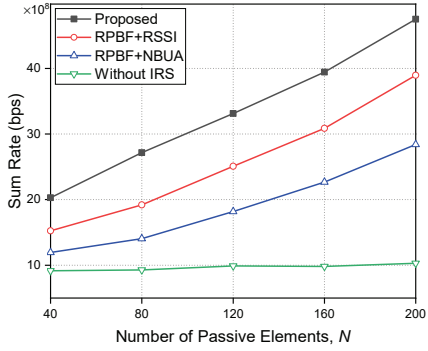}
  \caption{Sum rate versus $N$ with $M=30$.}
  \label{fig:IRS_N}
\end{figure}
\begin{figure}[t]
  \centering
  \includegraphics[width=5.6cm]{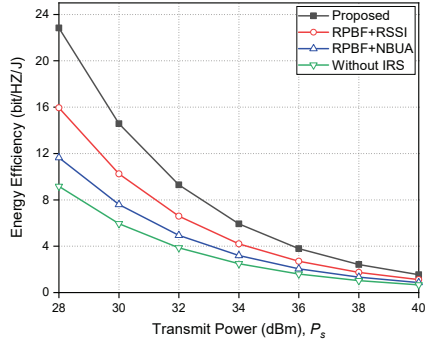}
  \caption{Energy efficiency versus $P_s$ with $M=30$, $N=60$.}
  \label{fig:PT_EE}
\end{figure}

Fig.\ref{fig:M} shows the impact of the number of antennas. In the figure, when the number of antennas increases at each BS, the sum rate increases and the proposed joint optimization algorithm gets the maximum sum rate. The reason is that, as the number of antennas increases, the number of antennas available for beamforming increase as well, which implies active beamforming becomes more efficient and results in the achievable rate of users to be increased. In addition, the proposed algorithm achieves significant gain in the sum rate of all users due to better passive beamforming gain and user association gain compared to other algorithms.

Besides, we evaluate the sum rate versus the number of passive elements $N$ in Fig. \ref{fig:IRS_N}. In the figure, the proposed algorithm outperforms other compared algorithms mainly due to the optimal gain of passive beamforming at IRS. Moreover, it can be seen that all these curves except the base line ascend as $N$ increases. This is because, when there are more passive elements at IRS, the more signals are reflected, the more effective the passive beamforming can be achieved.

Fig. \ref{fig:PT_EE} presents the energy efficiency under different settings of transmit power. In this simulation, when the total power is equal, our proposed algorithm can obtain the best energy efficiency than other algorithms because of better passive beamforming gain and user association gain. What's more, with increasing transmit power, the energy efficiency of all these algorithms reduces greatly, which implies that the system performance cannot be infinitely improved by increasing transmit power.

\section{Conclusion}
In this paper, we designed an IRS-assisted multiuser communication system to realize high robust and cost-effective mmWave communication. To improve the performance of the designed system, we formulated a sum rate maximization problem by jointly optimizing passive beamforming at IRS and user association. To solve this non-convex problem, an efficient alternating optimization algorithm is proposed. The simulation results show the sum rate promotion of our proposed algorithm under various scenarios. In particular, our proposed algorithm is able to provide up to 400\% higher sum rate and 200\% higher energy efficiency compared with the benchmark.
\section*{Acknowledgments}
This work was supported in part by the National Science Foundation of China (NSFC) (Grants 61631017, 91538203, U19B2044).
\vspace{-0.5em}

\bibliographystyle{ieeetr}
\bibliography{bibfile}

\end{document}